\documentclass[aps,prc,twocolumn,superscriptaddress,showpacs]{revtex4-1}

\usepackage{graphicx}
\usepackage{dcolumn}
\usepackage{bm}
\usepackage{ulem}
\usepackage{color}

\newcommand{\al}{$\alpha$}

\newcommand{\raX}{($\alpha$,$X$)}

\newcommand{\rag}{($\alpha$,$\gamma$)}
\newcommand{\ran}{($\alpha$,n)}

\newcommand{\rap}{($\alpha$,p)}

\newcommand{\sred}{$\sigma_{\rm{red}}$}
\newcommand{\ered}{$E_{\rm{red}}$}

\newcommand{\siii}{$^{33}$S}
\newcommand{\clvi}{$^{36}$Cl}
\newcommand{\arvi}{$^{36}$Ar}
\newcommand{\arvii}{$^{37}$Ar}
\newcommand{\sfact}{S-factor}

\begin{document}

\title{
   Unexpected properties of the $^{33}$S($\alpha$,p)$^{36}$Cl reaction cross
section at low energies
}

\author{Peter Mohr}
\email{WidmaierMohr@t-online.de}
\affiliation{
Diakonie-Klinikum, D-74523 Schw\"abisch Hall, Germany}
\affiliation{
Institute of Nuclear Research (ATOMKI), H-4001 Debrecen, Hungary}

\date{\today}

\begin{abstract}
New experimental data for the $^{33}$S($\alpha$,p)$^{36}$Cl reaction show a
very unusual energy dependence. Contrary to common findings for many other
$\alpha$-induced
reactions, statistical model calculations underestimate the measured cross
sections at very low energies. The relatively huge cross sections at these low
energies require a significant amount of single-particle strength in the
measured energy range which exceeds by far 100\,\% as soon as the additional
strength from the competing $^{33}$S($\alpha$,n)$^{36}$Ar reaction is taken
into account. In addition, the new data deviate from a general trend for the
energy dependence of $\alpha$-induced reaction cross sections.
\end{abstract}


\maketitle

Recently Bowers {\it{et al.}}\ \cite{Bow13} have measured the cross section
of the \siii \rap \clvi\ reaction at energies around the Coulomb barrier from
about 2.5 to 9\,MeV. The experiment has been done at the University of Notre
Dame using the activation technique. A \siii\ beam in combination with a
$^4$He gas cell was used for the production of the \clvi\ nuclei in inverse
kinematics, and the number of produced \clvi\ nuclei was determined from
accelerator mass spectrometry. The focus of \cite{Bow13} was the production of
short-lived 
radionuclides like \clvi\ in the early solar system by solar energetic
particles. The
main result of \cite{Bow13} was that the production of \clvi\ depends both on
the astrophysical event parameters and on the reaction cross sections of
several nuclear reactions, especially $^{34}$S($^{3}$He,p)\clvi , $^{34}$S(\al
,pn)\clvi , and \siii \rap \clvi . Additionally it was noticed that
calculations in the statistical model underpredict the experimental data
significantly, and it was not possible to find a parameter set which is able
to reproduce the new experimental data. 

It is the focus of the present study
to highlight and critically discuss the very unexpected properties of the new
experimental data for the \siii \rap \clvi\ reaction in \cite{Bow13}. For this
purpose I estimate an upper limit for the reaction cross section which is
derived from the single-particle strength, and I compare the data to a
systematics for so-called reduced cross sections and reduced energies.

\begin{figure}[htb]
\includegraphics[width=0.95\columnwidth,clip=]{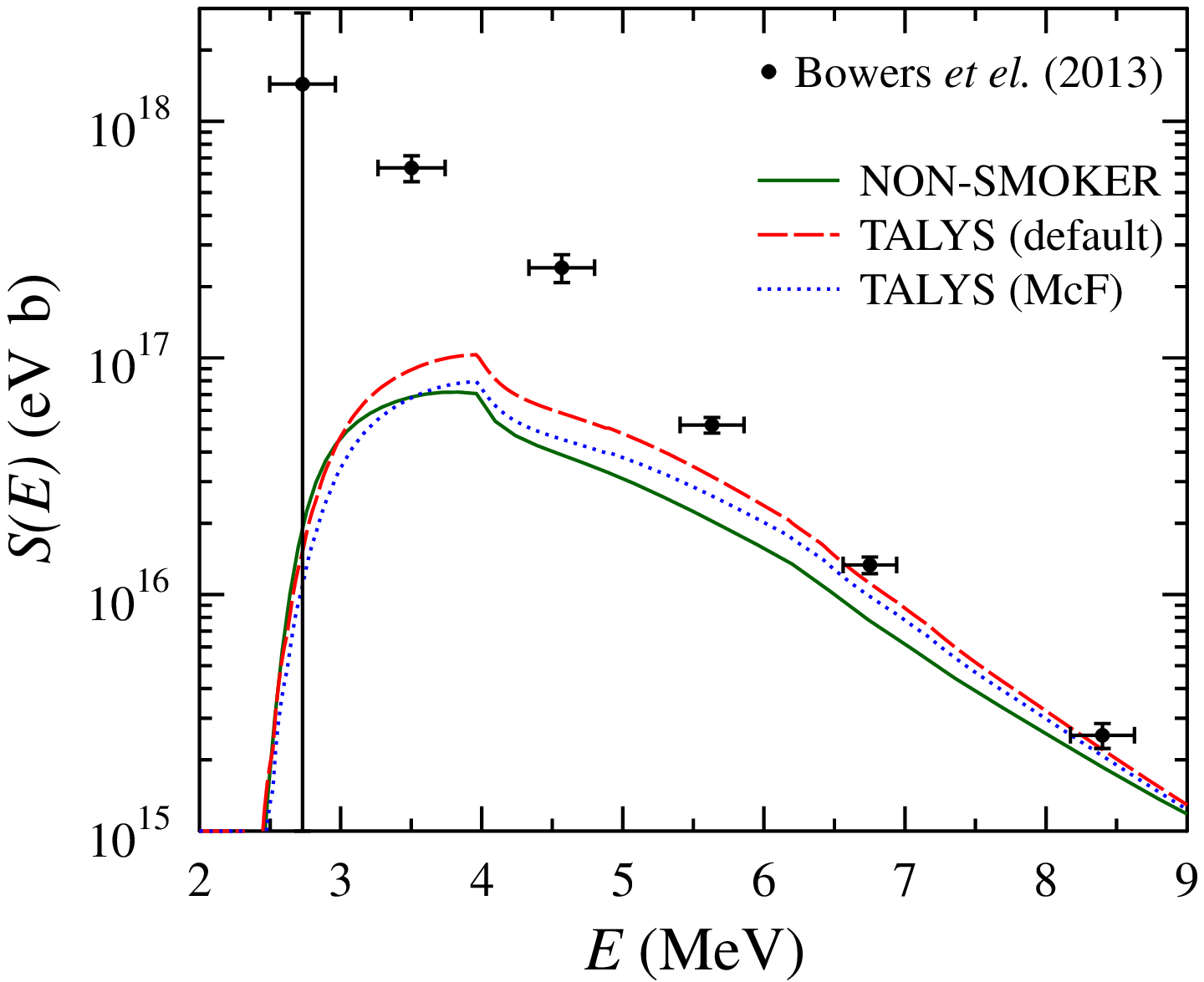}
\caption{
\label{fig:sfact}
(Color online)
Astrophysical \sfact\ of the \siii \rap \clvi\ reaction: comparison of the new
experimental data\footnote{Similar to Fig.~8 of \cite{Bow13}, the upper limit
  $\sigma < 0.1$\,mb is presented as $0.1 \pm 0.1$\,mb or $S(E) = (1.44 \pm
  1.44) \times 10^{18}$\,eV\,b.} 
\cite{Bow13} to statistical model calculations using the
NON-SMOKER default parameters and TALYS default and TALYS with the
McFadden/Satchler \al -nucleus potential \cite{McF66}.
}
\end{figure}

For many \al -induced reaction cross sections at energies around the Coulomb
barrier it is found that typical statistical model calculations are able to
reproduce the experimental data above the Coulomb barrier whereas below the
barrier experimental cross sections are significantly overestimated, see
e.g.\ \cite{Som98,Gyu06,Ozk07,Rap08,Cat08,Yal09,Gyu10,Kis11,Kis12,Sau11,Mohr11,Kis13}.
Similar to this general finding, the present \siii \rap \clvi\ data
\cite{Bow13} are roughly reproduced at the highest energy. However, in strict
contrast to the typical behavior, at lower energies the experimental data 
are not overestimated, but
dramatically underestimated. In Fig.~\ref{fig:sfact} the experimental data
\cite{Bow13} are compared to the results of two widely used statistical model
codes TALYS \cite{TALYS} and NON-SMOKER \cite{NS} (results
taken from \cite{RauWEB}). It has been stated that the typical overestimation
of cross sections at very low energies is related to the missing energy
dependence of the widely used \al -nucleus potential of McFadden and Satchler
\cite{McF66}, and various attempts have been made to provide \al -nucleus
potentials with an energy-dependent imaginary part
\cite{Som98,Sau11,Mohr11,Mohr13,Avr10,Dem02} which essentially 
defines the reaction cross section. A reduction of reaction cross sections at
low energies is related to an imaginary potential which decreases towards
lower energies; such a behavior can be naturally understood from the smaller
number of 
open reaction channels at low energies. However, the reproduction of the new
\siii \rap \clvi\ data requires an increasing imaginary part at low energies
which is at least an unexpected result.

Contrary to the original presentation in
Fig.~8 of \cite{Bow13} where the cross section $\sigma(E)$ is plotted versus
the energy per nucleon of the \siii\ projectile, here I present the
astrophysical \sfact\ $S(E)$ versus the center-of-mass energy $E$. (Note that
$E$ refers to the center-of-mass energy in this paper except explicitly
stated.) The \sfact\ vs.\ $E$ presentation should show only a mild energy
dependence because the dominating Coulomb effect on the energy dependence is
separated from the data.

\begin{figure}[b]
\includegraphics[width=0.95\columnwidth,clip=]{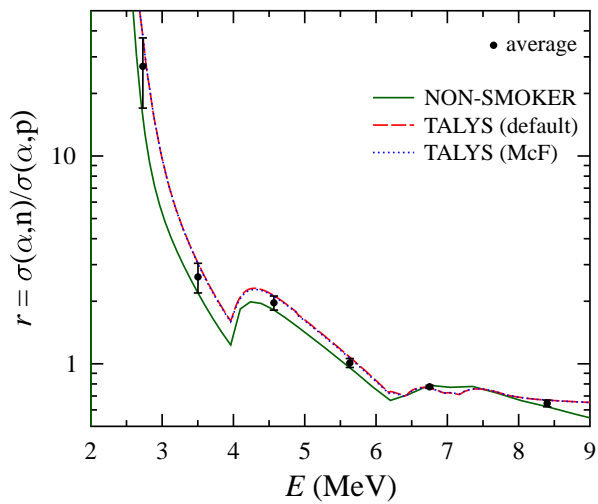}
\caption{
\label{fig:ratio}
(Color online)
Calculated ratio $r = \sigma(\alpha,n)/\sigma(\alpha,p)$. The data points are
avarage values from the statistical model calculations to estimate the
\siii \ran \arvi\ cross section from the experimental \siii \rap \clvi\ data
\cite{Bow13}.
}
\end{figure}

The \siii \rap \clvi\ reaction has a slightly negative $Q$-value of $Q =
-1.93$\,MeV. Almost the same $Q$-value is found for the \siii \ran
\arvi\ reaction with $Q = -2.00$\,MeV. Thus, it can be expected that close
above the almost common threshold the \ran\ cross section dominates because
the proton emission from the \arvii\ compound nucleus
is suppressed by the Coulomb barrier. At higher energies
the \rap\ and \ran\ reactions have similar cross sections, even slightly
favoring the \rap\ reaction because of the larger number of open channels for
the decay of \arvii\ into the residual odd-odd
\clvi\ nucleus. The calculated ratio $r = \sigma(\alpha,n)/\sigma(\alpha,p)$
is shown in Fig.~\ref{fig:ratio}. 

Calculations of the \siii \rap \clvi\ reaction cross section in \cite{Bow13}
and in this work are based on the statistical model (StM). This requires a
brief discussion on the applicability of the StM and on the relevance 
of the various ingredients of the StM. A basic prerequisite of the StM
is a sufficiently high level density in the compound nucleus (here
\arvii ). The level density increases with excitation energy and with the mass
number $A$ of the compound nucleus. The new data of \cite{Bow13} correspond to
$E^\ast \approx 9 - 16$\,MeV in the even-odd nucleus $^{37}$Ar. The smooth
energy dependence of the experimental \sfact s (see Fig.~\ref{fig:sfact})
within the relatively broad energy widths of the data points ($\approx 380 -
480$\,keV) is a clear experimental indication that the measured cross
sections are composed of many overlapping resonances and not of very few
isolated resonances; thus, for the measured average cross sections within the
experimental energy windows the StM should be applicable. This finding is
further strengthened by the successful application of the StM to various
reactions in this mass region (e.g., \al -induced reactions for target nuclei
between $^{27}$Al and $^{51}$V, see \cite{How74,Von83,Son00}). 

The calculations in the StM obviously depend on the chosen input
parameters which are the level densities, optical potentials, and $\gamma$-ray
strength functions. Fortunately, in the present case the relevance of the
various ingredients can be nicely disentangled. The cross section of a
\raX\ reaction in the StM is proportional to $(T_{\alpha,0} T_X)/T_{\rm{total}}$
with the transmissions $T_i$ as e.g.\ defined in \cite{Rau11} and
the total transmission $T_{\rm{total}} = \sum_i T_i$. Except very close above
the respective thresholds around $E \approx 2$\,MeV, the transmissions $T_n$
and $T_p$ into the \arvi -n and \clvi -p channels are much larger than
$T_\alpha$ and $T_\gamma$. Hence, $T_{\rm{total}} \approx T_n + T_p$ which
leads to the following approximate proportionalities: 
$\sigma$\ran\ $\sim (T_{\alpha,0} T_n)/(T_n+T_p)$,
$\sigma$\rap\ $\sim (T_{\alpha,0} T_p)/(T_n+T_p)$, 
$\sigma$\ran /$\sigma$\rap\ $\sim T_n/T_p$, 
and
$\sigma_{\rm{total}} \sim T_{\alpha,0}$.
Consequently, the total reaction cross section is practically only sensitive
to the chosen \al -nucleus potential, and the ratio $r =
\sigma(\alpha,n)/\sigma(\alpha,p)$ depends only on the chosen nucleon-nucleus
potential. The parametrization of the level densities has almost no effect on
the calculated cross sections because the StM codes \cite{TALYS,NS} use
the well-known experimental level schemes at low energies for the involved
nuclei close to stability. These low-lying states are the most important final
states at low energies, see e.g.\ Eq.~(64) in \cite{Rau11}. 

As expected, the calculated ratio $r$ is practically insensitive to the choice
of the \al -nucleus potential (see Fig.~\ref{fig:ratio}); thus, such ratios are 
an excellent test for the remaining parameters of the StM except
the \al -nucleus potential. The ratio $r$ is nicely constrained because of the
good agreement of the different calculations using a variety of global
nucleus-nucleon potentials; in particular, the potentials of
Refs.~\cite{Kon03,Bau01,Jeu77} have been used for the calculations in
Fig.~\ref{fig:ratio}. It should be noted that in any case the same
nucleon-nucleus potential has been used for the proton and the neutron
channel; this leads to a relatively well-constrained ratio $r$. No attempt was
made using inconsistent combinations like the potential from \cite{Kon03} for
the neutron channel and from \cite{Jeu77} for the proton channel.

Summarizing the above arguments, the StM should be well applicable to the
\siii \rap \clvi\ reaction under study. The huge discrepancy between the StM
calculations and the recent experimental data \cite{Bow13} at low energies
cannot be explained in a simple way. Thus, further investigations of the
unexpected energy dependence of the \siii \rap \clvi\ reaction cross sections
have been performed.

At the energies under study in \cite{Bow13}, the \siii \rap \clvi\ and \siii
\ran \arvi\ reactions are governed by many resonances corresponding to levels
in the \arvii\ compound nucleus at excitation energies from about 9 to
16\,MeV. Following e.g.\ the definitions in \cite{Ili97}, the decay width
$\Gamma_{X,L}$ for a particular channel $X$ with relative
angular momentum $L$ can be estimated from the single-particle (s.p.) limit
\begin{equation}
\Gamma^{s.p.}_{X,L}(E) = \frac{ 2 k R }{F_L^2(E,R) + G_L^2(E,R)} \, \times \,
\frac{\hbar^2}{\mu R^2}
\label{eq:WL}
\end{equation}
with the wave number $k$, the nuclear radius $R = R_0 \times (A_P^{1/3} +
A_T^{1/3})$ and $R_0 = 1.25$\,fm, the regular and irregular Coulomb functions
$F_L$ and $G_L$, and the reduced mass $\mu$. The decay widths with small
angular momenta $L$ are dominating because of the smaller centrifugal barrier
and because all $J^\pi$ of the relevant levels
in \siii , \clvi , \arvi , and \arvii\ are small. Furthermore it is found that
the larger Coulomb barrier in the \al\ channel leads to $\Gamma_\alpha \ll
\Gamma_p$ and $\Gamma_\alpha \ll \Gamma_n$ for practically the full energy
range under study. Consequently, the total width $\Gamma$ is essentially
defined by the sum of the dominating partial widths $\Gamma_p$ and $\Gamma_n$
(the radiation width $\Gamma_\gamma$ is also small). The cross sections
$\sigma(E)$ measured in \cite{Bow13} are the average cross sections (averaged
over all contributing resonances within the broad experimental
energy window $\Delta E$
which is defined by the energy loss of the \siii\ beam in the helium gas
target). From a detailed study of partial widths \cite{Pog13} it has been
concluded that on average experimental $\Gamma_\alpha$ are about 2\,\% of the
single-particle limit in Eq.~(\ref{eq:WL}).

\begin{figure}[htb]
\includegraphics[width=0.95\columnwidth,clip=]{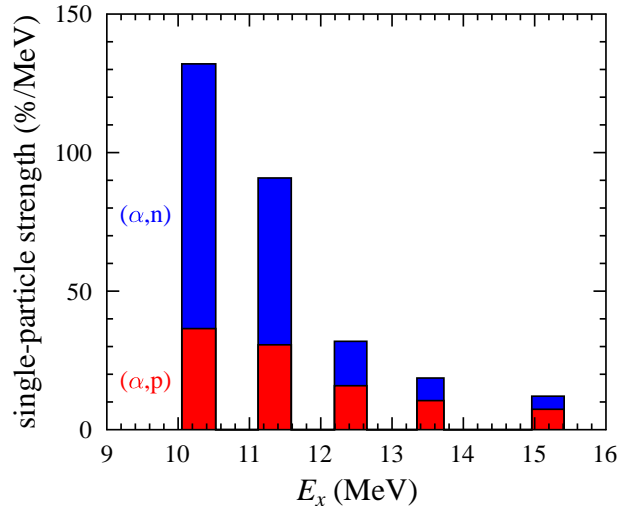}
\caption{
\label{fig:str}
(Color online)
Single-particle strength per MeV for the \siii \rap \clvi\ (red) and the \siii
\ran \arvi\ (blue) reactions vs.\ excitation energy $E_x$ in \arvii\ ($E_x = E
+ S_\alpha$ with the \al\ separation energy $S_\alpha = 6.787$\,MeV).
}
\end{figure}

From the single-particle limits of the decay widths $\Gamma^{s.p.}_{\alpha,L}$
it is possible to calculate a limit for the cross section
$\sigma^{s.p.}_L$ for the $L$-th partial wave in an energy interval $\Delta
E$. As the condition $\Gamma < \Delta E \approx 500$\,keV is fulfilled for
most states, this can be done easily using the 
narrow-resonance formalism (using the resonance strength $\omega \gamma
\approx \omega \Gamma_\alpha$). For an isolated resonance with
$\Gamma_{\alpha,L}^{s.p.}$ this leads to
\begin{equation}
\sigma^{s.p.}_L = \frac{1}{\Delta E} \int \sigma(E) dE \approx 
\frac{\pi^2 \hbar^2}
     {\mu E \, \Delta E} \, \times \, (2L+1) \, \Gamma^{s.p.}_{\alpha,L}
\label{eq:av}
\end{equation}
with the experimental energy range $\Delta E$. The results are listed in Table
\ref{tab:av}. Let me explain e.g.\ the second line of Table \ref{tab:av} in
more detail. If a $s$-wave resonance ($L = 0$; $J^\pi = 3/2^+$) with full
single-particle strength ($\theta^2_\alpha = 1.0$) is located within the
energy window $\Delta E = 0.476$\,MeV around $E = 3.503$\,MeV (i.e., between
3.265 and 3.741\,MeV), it will contribute with $\sigma^{s.p.}_{\alpha,L=0} =
2.47$\,mb to the measured average cross section at this energy. The same
result is obtained if several $s$-wave resonances with a summed
$\theta^2_\alpha = 1.0$ are located within this energy window (neglecting the
energy dependence of the width $\Gamma^{s.p.}_{\alpha,L}$ within the energy
window $\Delta E$; see also below). For $p$-waves ($L = 1$, $J^\pi = 1/2^-,
3/2^-, 5/2^-$) the single-particle cross section within $\Delta E$ is
$\sigma^{s.p.}_{\alpha,L=1} = 5.03$\,mb.  The sum over single-particle cross
sections from all partial waves is 13.8\,mb. If the experimental result were
also 13.8\,mb, then all single-particle strength for all partial waves must
be located in this energy window, thus no \al\ strength remaining for energies
outside $\Delta E$. The experimental result of $2.4 \pm 0.3$\,mb corresponds
to 17\,\% of the single-particle strength of all partial waves which is still
a very high value for a $\Delta E \approx 0.5$\,MeV energy window. Summing up
the \al -strength over the five measured intervals $\Delta E$ reaches already
45\,\% 
of the single-particle strength for all partial waves, and assuming a smooth
energy dependence for the \al -strength between the data points leads to about
100\,\% of the available single-particle strength. 
The \al -strength is shown in Fig.~\ref{fig:str}; here the \al -strength has
been normalized to a 1\,MeV interval because the energy intervals of the
experimental data points are not exactly identical. Furthermore, it can be
read from Fig.~\ref{fig:str} that the single-particle strength increases
strongly towards lower energies. 
\begin{table*}[htb]
\caption{\label{tab:av}
Cross section $\sigma^{s.p.}_L$ (in mb) within the experimental energy range
$\Delta E$ from the single-particle limits $\Gamma^{s.p.}_{\alpha,L}$ in
Eq.~(\ref{eq:WL}). The contribution of partial waves with $L > 6$ to the
sum $\sum \sigma^{s.p.}_L$ remains negligible ($\lesssim 1$\,\%).
}
\begin{center}
\begin{tabular}{cccccccccccc}
\multicolumn{1}{c}{$E$ (MeV)} 
& \multicolumn{1}{c}{$\Delta E$ (MeV)}
& \multicolumn{1}{c}{$L=0$}
& \multicolumn{1}{c}{$L=1$}
& \multicolumn{1}{c}{$L=2$}
& \multicolumn{1}{c}{$L=3$}
& \multicolumn{1}{c}{$L=4$}
& \multicolumn{1}{c}{$L=5$}
& \multicolumn{1}{c}{$L=6$}
& \multicolumn{1}{c}{$\sum \sigma^{s.p.}_L$}
& \multicolumn{1}{c}{$\sigma_{\rm{exp}}$}
& \multicolumn{1}{c}{$\sigma_{\rm{exp}}/\sum \sigma^{s.p.}_L$} \\
\hline
2.730 & 0.465 & 
0.09  & 0.17  & 0.12  & 0.05  & 0.01  & 0.00   & 0.00 &
0.44  & $<0.1$ & $<0.23$ \\
3.503 & 0.476 & 
2.47  & 5.03  & 3.90  & 1.78  & 0.53  & 0.11   & 0.02 &
13.83 & 2.4(3) & 0.17 \\
4.568 & 0.465 & 
38.5  & 84.8  & 75.7  & 41.2  & 15.1  & 3.9   & 0.7 &
259.9 & 37(5) & 0.14 \\
5.632 & 0.454 & 
167.6  & 403.6  & 423.2  & 282.7  & 128.9  & 41.7   & 9.9 &
1458 & 105(8) & 0.07 \\
6.751 & 0.378 & 
426.3  & 1109.7  & 1356.1  & 1121.4  & 651.9  & 268.2   & 79.9 &
5014 & 199(16) & 0.04 \\
8.400 & 0.454 & 
565.0  & 1568.2  & 2200.2  & 2275.6  & 1791.7  & 1041.5   & 435.9 &
9878 & 330(40) & 0.03 \\
\hline
\end{tabular}
\end{center}
\end{table*}

As already stated above, the cross section for the \siii \ran \arvi\ reaction
exceeds the \siii \rap \clvi\ cross section at low energies (see also
Fig.~\ref{fig:ratio}), and therefore the measured \al -strength in the
\rap\ reaction is only a small part of the total \al -strength. From the
calculated ratio $r = \sigma(\alpha,n)/\sigma(\alpha,p)$
in Fig.~\ref{fig:ratio} the cross section of
the \siii \ran \arvi\ reaction can be derived, and the corresponding \al
-strength in the \ran\ reaction can be estimated (see Fig.~\ref{fig:str}). The
total \al -strength exceeds by far the single-particle limit; this is again a
very unexpected finding.

\begin{figure}[htb]
\includegraphics[width=0.95\columnwidth,clip=]{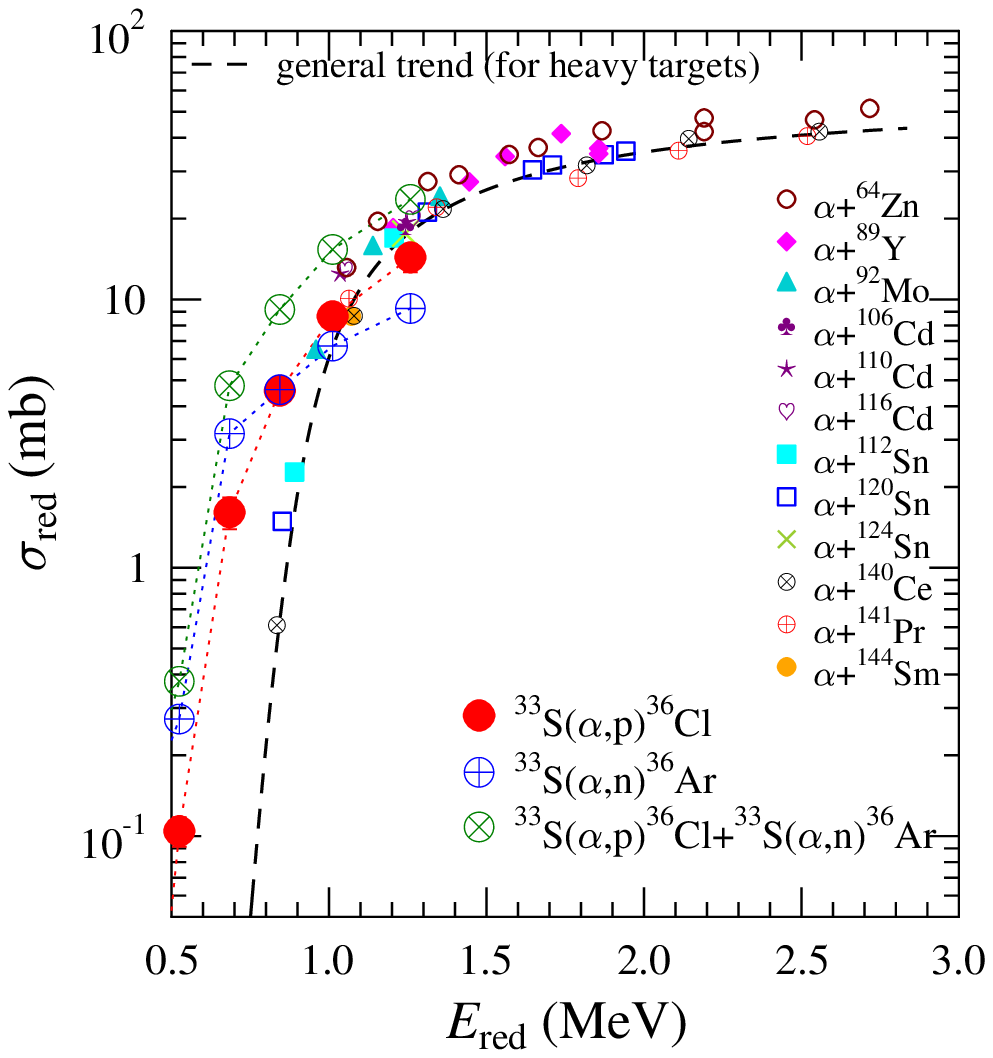}
\caption{
\label{fig:sred}
(Color online)
Total reduced cross section \sred\ versus reduced energy \ered\ for many \al
-induced reactions \cite{Mohr13}, compared to the recent \siii \rap
\clvi\ \cite{Bow13} and \siii \ran \arvi\ cross sections; the latter is
calculated from the ratio $r$ in Fig.~\ref{fig:ratio}.
}
\end{figure}

A further comparison can be made in the following way. A common behavior for
the total reaction cross section of \al -induced reactions for various targets
at different energies is found when so-called reduced cross sections
\sred\ are plotted versus reduced energies \ered\ \cite{Mohr13}:
\begin{eqnarray}
\sigma_{\rm{red}} & = & \frac{\sigma}{(A_P^{1/3}+A_T^{1/3})^2}
\label{eq:sred} \\
E_{\rm{red}} &= & \frac{(A_P^{1/3}+A_T^{1/3})}{Z_P Z_T} \times E
\label{eq:ered}
\end{eqnarray}
Deviations from this common behavior have been found e.g.\ for weakly bound
projectiles like $^6$He \cite{Far10}. In Fig.~\ref{fig:sred} total reaction
cross sections are shown for many target nuclei, and the dashed line shows the
general trend (calculated from the new ATOMKI-V1 potential \cite{Mohr13} for
$^{140}$Ce \cite{Mohr13b}). The cross section of the \siii \rap
\clvi\ reaction at the highest energy is found below the general trend, but as
soon as the additional contribution of the \siii \ran \arvi\ reaction is taken
into account, the total reaction cross section of \siii\ fits nicely into the
general trend of the data. However, the energy dependence of the \siii \rap
\clvi\ reaction is much flatter than the usual trend, and even the cross
sections of the \siii \rap \clvi\ reaction at lower energies by far exceed the
general trend of total reaction cross sections. Adding the additional
contribution of the \siii \ran \arvi\ reaction obviously sharpens this
discrepancy. 

The total reaction cross sections which are presented as \sred\ in
Fig.~\ref{fig:sred} have been derived from the analysis of elastic scattering
angular distributions. Such experiments are difficult for small \ered\ in the
mass range of the present study because the angular distributions may be
dramatically affected by compound resonances. Nevertheless, total reaction
cross sections can be estimated from the contributions of the different open
channels. If one considers the sum of measured \rap , \ran , and \rag\ cross
sections as total reaction cross section, then the data in \cite{Gyu12} for
$^{64}$Zn can be used to extend the \sred\ values from elastic scattering
down to lower energies, and e.g.\ at \ered\ $ = 0.7$\,MeV a reduced cross
section of \sred\ $ \approx 0.18$\,mb is found. This value for $^{64}$Zn is
slightly higher than the general trend for heavier nuclei but still an order
of magnitude below the results for \siii .

Recently, a new systematics has been suggested for the comparison of cross
sections from various fusion reactions. This new systematics uses a different
way for the calculation of the reduced energy and takes into account the
$Q$-value of the fusion reaction \cite{Wol13}. A similar deviation of the new
data for \siii\ from the general trend for heavier nuclei is also found in
this case.

In the present study the energy dependence of the widths
$\Gamma^{s.p.}_{\alpha,L}$ within the energy window $\Delta E$ is
neglected. Because a similar energy dependence of the cross section
$\sigma(E)$ within $\Delta E$ has also been neglected in \cite{Bow13}, the
derived strengths from the ratio $\sigma_{\rm{exp}}/\sum_L \sigma^{s.p.}_L$
should be consistent. Taking into account this energy dependence in the analysis
of the experimental data will lead to slightly reduced cross sections;
however, the changes should be of the order of 10\,\% and thus do not affect
the main conclusion of the present study.

In summary, the new experimental data for the \siii \rap \clvi\ reaction
\cite{Bow13} show a very unexpected behavior in all respects. First of all,
the experimental \sfact\ data increase dramatically towards lower energies and
exceed any theoretical prediction within the statistical model whereas usually
the calculated cross sections exceed the experimental results. The huge
\sfact\ of the \siii \rap \clvi\ reaction at low energies requires a
significant amount of \al -strength already within the measured energy
intervals. It reaches already 100\,\% of the single-particle strength if one
assumes a smooth energy dependence of the \al -strength between the measured
energy intervals. The excess of \al -strength far over 100\,\%
of the single-particle strength  becomes visible
as soon as the additional strength in the \siii \ran \arvi\ reaction is taken
into account. Finally, the new experimental data deviate significantly from
the general behavior of the energy dependence of \al -induced reaction cross
sections in the \sred\ vs.\ \ered\ presentation. Hence, the new experimental
data for the \siii \rap \clvi\ reaction -- if confirmed -- 
point to an extraordinary behavior of the \siii -\al\ system and
will be an extreme
challenge for theoretical models.

\smallskip
This work was supported by OTKA (K101328 and K108459).

\bigskip
\noindent
{\bf\underline{Note added in proof:}}\\
Very recently, new data for the $^{23}$Na\rap $^{26}$Mg reaction have been
measured \cite{Alm14}. The analysis of these data shows that statistical model
calculations underestimate the experimental data by about a factor of 40,
i.e.\ even more dramatic than for the \siii \rap \clvi\ reaction studied in
this work. Further investigations of \rap\ reactions in the mass range $20
\lesssim A \lesssim 50$ are urgently needed to resolve these unexpected huge
discrepancies.

\end{document}